\documentclass[11pt]{article}
\usepackage{graphicx,floatflt,amssymb,epsfig,epsf,subfigure} 
\def\be {\begin{equation}}
\def\ee {\end{equation}}
\def\bea {\begin{eqnarray}}
\def\eea {\end{eqnarray}}
\def\bfl {\begin{flushleft}}
\def\efl {\end{flushleft}}

\def\b {\boldmath}

\def\al {\alpha}
\def\bt {\beta}                                                               \def\s {\sigma}
\def\g {\gamma}

\def\l {\lambda}
\def\L {\Lambda}

\def\bra{\langle}
\def\ket{\rangle}

\def\opcit(#1){ {\em op. cit.}, #1}
\def\etal {\em et al.}

\def\issue(#1,#2,#3){#1 (#3) #2} 
\def\APP(#1,#2,#3){Acta Phys.\ Polon.\ \issue(#1,#2,#3)}
\def\ARNPS(#1,#2,#3){Ann.\ Rev.\ Nucl.\ Part.\ Sci.\ \issue(#1,#2,#3)}
\def\CPC(#1,#2,#3){Comp.\ Phys.\ Comm.\ \issue(#1,#2,#3)}
\def\CIP(#1,#2,#3){Comput.\ Phys.\ \issue(#1,#2,#3)}
\def\EPJC(#1,#2,#3){Eur.\ Phys.\ J.\ C\ \issue(#1,#2,#3)}
\def\EPJD(#1,#2,#3){Eur.\ Phys.\ J. Direct\ C\ \issue(#1,#2,#3)}
\def\IEEETNS(#1,#2,#3){IEEE Trans.\ Nucl.\ Sci.\ \issue(#1,#2,#3)}
\def\IJMP(#1,#2,#3){Int.\ J.\ Mod.\ Phys. \issue(#1,#2,#3)}
\def\JHEP(#1,#2,#3){J.\ High Energy Physics \issue(#1,#2,#3)}
\def\JPG(#1,#2,#3){J.\ Phys.\ G \issue(#1,#2,#3)}
\def\MPL(#1,#2,#3){Mod.\ Phys.\ Lett.\ \issue(#1,#2,#3)}
\def\NP(#1,#2,#3){Nucl.\ Phys.\ \issue(#1,#2,#3)}
\def\NIM(#1,#2,#3){Nucl.\ Instrum.\ Meth.\ \issue(#1,#2,#3)}
\def\PL(#1,#2,#3){Phys.\ Lett.\ \issue(#1,#2,#3)}
\def\PRD(#1,#2,#3){Phys.\ Rev.\ D \issue(#1,#2,#3)}
\def\PRL(#1,#2,#3){Phys.\ Rev.\ Lett.\ \issue(#1,#2,#3)}
\def\PTP(#1,#2,#3){Progs.\ Theo.\ Phys. \ \issue(#1,#2,#3)}
\def\RMP(#1,#2,#3){Rev.\ Mod.\ Phys.\ \issue(#1,#2,#3)}
\def\SJNP(#1,#2,#3){Sov.\ J. Nucl.\ Phys.\ \issue(#1,#2,#3)}
\def\ZPC(#1,#2,#3){Zeit.\ Phys.\ C \issue(#1,#2,#3)}

\begin{document} 
\begin{flushright} 
IMSc-PHYSICS/08-2005\\
\end{flushright} 
\vskip 30pt 
 
\begin{center} 
{\Large Determination of \boldmath $\gamma$ from the \boldmath 
$\eta - \eta'$ mixing within QCD Factorization}\\
\vspace*{1cm} 
\renewcommand{\thefootnote}{\fnsymbol{footnote}} 
{\large {\sf Prasanta Kumar Das}~\footnote{E-mail: dasp@imsc.res.in} }\\ 
\vspace{10pt} 
{\small The Institute of Mathematical Sciences, \\
C.I.T Campus, Taramani, Chennai-600113, India.}
 
\normalsize 
\end{center} 

\vskip1cm \small
\begin{abstract} 
The charmless nonleptonic $B\to \eta K$ decay is an useful probe to test the strong 
interaction dynamics part of the Standard Model. Within the QCD factorization
 framework, we analyse this particular decay by using the most recent data of 
$BR(B^+ \to \eta K^+)$ and $A_{cp}(B^+ \to \eta K^+)$ available in the 
Heavy Flavour Averaging Group (HFAG) webcite. Using these data we constraint the 
unitarity angle $\gamma$ and $\eta-\eta'$ mixing angle $\theta$. We find that such 
constraint is scale dependent, e.g. for $\gamma = 70^o$, the 
data for $BR(B^+ \to \eta K^+)$ (=~$(2.6 \pm 0.5)\times 10^{-6}$) 
suggests that $\theta$ should lie in between $-46^o$ and $-44^o$, 
$-30^o$ and $-26^o$ for $\mu = m_b/2$. For $\mu = m_b$, the same $BR$ data (with 
the same $\gamma$) suggests that $\theta$ should lie in between $-54^o$ 
and $-50^o$, $-30^o$ and $-26^o$. The allowed region followed from 
$A_{cp}=-0.25\pm 0.14$, is found to be rather wider than that obtained from the $BR(B^+ \to \eta K^+)$ data. For $\gamma=70^o$ and $\theta = -21.3^o$,
we find $BR(B^-(+) \to \eta K^-(+))= 2.93(4.91) \times 10^{-6}$ and 
$A_{cp}=-0.252$ at $\mu=m_b$, and about $ 2.98(5.72) \times 10^{-6}$ and $-0.315$
 corresponding to $\mu=m_b/2$. We investigate the role of the power corrections  
in such constraints.  
\end{abstract}

\vspace{10pt}
\bfl
\vspace*{0.05in}
\efl


\newpage
\section{Intrduction} \label{sec:section1}
The nonleptonic weak hadronic decays of the $B$ meson 
is an useful probe to test the Standard Model(SM), particularly the dynamics of it's strong
interaction (QCD) part. 
Due to the non-perturbative features, the amplitudes for such hadronic $B$-decays,
are difficult to calculate directly from the
QCD lagrangian. For the phenomenological study, several
factorization hypotheses, like naive-factorization(NF), generalized
factorization(GF) and QCD factorization(QCDF) were introduced and
has been quite successful in explaining  the data for
exclusive $B$ decays, like $B \rightarrow PP,~PV,~VV,~PT,~VT$
(where $P$, $V$ and $T$ stands for pseudoscalar, vector and
tensor meson) reported by the CLEO, BABAR and BELLE
collaborations \cite{Anast,Aubert:2003xc,Zhang:2003up}. Out of all these hypothesis, 
the QCD factorization hypothesis \cite{qcdf:BBNS} particularly, has become very much 
popular in recent years 
because of it's successful explanation of the several $B$ factories data.  
Studying the nonleptonic $B^{-(+)} \to \eta K^{-(+)}$ decay within the QCD factorization, exploring 
it's CP-violation aspects, is a well motivated topic and is the main concern of the present work.

The only source of the CP violation in the Standard Model(SM) is the CKM matrix 
$V_{CKM}$ \cite{KM}, which is arised  due to the misalignment
of the mass and weak interaction eigenstates. It is expressed via the charge 
current-current interaction lagrangian ${\cal L}^{CC}_{int}$, defined by
\bea 
{\cal L}^{CC}_{int}=-\frac{g}{\sqrt{2}}\left(\begin{array}{ccc}
\bar u_L & \bar c_L & \bar t_L \end{array}\right) \gamma_\mu
V_{CKM}\left( \begin{array}{c} d_L\\s_L\\b_L \end{array}\right)
W_\mu^+ +h.c. 
\eea
 In the Wolfenstein parametrization $V_{CKM}$ is given by
\bea \label{eqn:ckm}
V_{CKM}=\left(\begin{array}{ccc} V_{ud} & V_{us} & V_{ub} \\
 V_{cd} & V_{cs} & V_{cb} \\
 V_{td} & V_{ts} & V_{tb} \\
\end{array} \right)=
\left(\begin{array}{ccc}
1-\frac 12 \lambda^2 & \lambda & A\lambda^3 (\rho -i\eta )\\
-\lambda & 1-\frac 12 \lambda^2 & A\lambda^2 \\
A\lambda^3 (1-\rho - i\eta ) & -A\lambda^2 & 1 \end{array} \right)
+{\cal {O}} (\lambda^4) ,
\eea
where $\lambda = V_{us} = sin\theta_c$, $\theta_c$, the Cabibbo mixing angle.
The angle $\gamma$  related to the phase of the CKM element $V_{ub}$ i.e. 
$\gamma = arg(V^*_{ub})$ (which follows from 
$V_{ub} = |V_{ub}| e^{-i \gamma} = A\lambda^3 (\rho -i\eta )$), 
amounts to CP violation in the SM. The angle 
$\gamma$ is one of the three angles $\al,~\bt$ and $\gamma$ of the unitarity traingle 
and any nonzero value of it (resulting into non zero area of the unitarity traingle) yields the CP 
violation in the SM. Hence, a precise measurement of the unitarity angle $\gamma$ can be possible from 
the CP violating transition, which is being considered in the present work.  There exist 
bounds on the angle 
$\gamma$ and one such bound is followed from the time-dependent CP asymmetry $S_{\pi\pi}$ measurement. 
The current average experimental value of the 
$S_{\pi\pi}$ which is about $ = -0.49 \pm 0.27$, yields 
\cite{qcdf1:BN}
\bea
\gamma = \left(66^{+19}_{-16}\right)^o ~~or~~ \left(174^{+9}_{-8}\right)^o,
\eea
where both set of solutions are consistent with each other and the first set 
is consistent with the unitarity-traingle fit.  Our next concern is the mixing angle $\theta$ between 
the psedoscalar $\eta$ and $\eta'$ mesons, a matter of great interst from the time when the 
$SU(3)$ flavour symmetry was proposed. In the simplest scenario where the $\eta$ and $\eta'$ mesons do 
not mix with other pseudoscalar mesons, like excited quarkonium states, gluonium or exotics, the $\eta$, $\eta'$ wave functions can be written as 
\bea
|\eta \ket &=& cos\theta ~|\eta_8\ket - sin\theta ~|\eta_0\ket, \\
|\eta'\ket &=& sin\theta ~|\eta_8\ket + cos\theta ~|\eta_0\ket,
\eea 
where the $SU(3)$ basis states $|\eta_8 \ket$ and $|\eta_0 \ket$ are given by
\bea
|\eta_8 \ket &=& \frac{1}{\sqrt{6}} | u \bar{u} + d \bar{d} - 2 s \bar{s} \ket, \\
|\eta_0 \ket &=& \frac{1}{\sqrt{3}} | u \bar{u} + d \bar{d} +   s \bar{s} \ket. 
\eea
From the accumulated data of the decay widths $\Gamma[\eta \to \gamma \gamma]=(0.46 \pm 0.04)$ KeV,~$\Gamma[\eta' \to \gamma \gamma]=(4.26 \pm 0.19)$ KeV and $\Gamma[\pi^0 \to \gamma \gamma]=(7.7 \pm 0.55)$ eV, one finds \cite{b2etap:Ali},
\bea
\theta = -21.3^o \pm 2.5^o.
\eea
In the present analysis we will treat the unitarity angle $\gamma$ and 
 mixing angle $\theta$ as the unknown parameters and 
use the most recent data for the direct CP asymmetry $A_{cp}(B^+\to \eta K^+)$ and $BR(B^+\to \eta K^+)$ 
\cite{HFAG05} to put constraints on them.

 The organization of the paper is as follows. Section \ref{sec:section2} begins with the general discussion of the nonleptonic decay of a $B$ meson within the
QCD factorization framework, which includes the discussion 
of several non-factorizable corrections, e.g. vertex, penguin, hard-spectator and 
the weak annihilation corrections. They are present in the QCD factorization 
 framework, but absent in the simplest naive factorization(NF) framework. We
obtain the SM decay amplitude ${\cal {M}}_{SM}(B^+\to \eta K^+)$, $BR(B^+\to \eta K^+)$ 
and $A_{cp}(B^+\to \eta K^+)$. Section \ref{sec:section3} is fully devoted to the numerical analysis. 
After describing several numerical inputs i.e. the CKM matrix elements, effective coefficients, 
quarks masss, decay constants, form factors, we discuss the constraints in 
the $\theta -\gamma$ contour plane, which is being obtained by using  
the most recent data of $BR(B^+\to \eta K^+)$ and $A_{cp}(B^+\to \eta K^+)$, available in the 
HFAG webcite \cite{HFAG05}. In Section \ref{sec:section4},
we summarize our results and made our conclusion.

\section{$B^- \rightarrow \eta ~K^-$ decay within the QCDF framework:} 
\label{sec:section2}

 The most general effective weak Hamiltonian  
${\cal H}_{\rm eff}^{\Delta B = 1}$ for the non-leptonic
$\Delta B = 1$ transitions can be expressed via the operator product expansion
(OPE) \cite{brev:bbl}
 \bea \label{eqn:effH}
\label{eff}
 {\cal H}_{\rm eff} = \frac{G_F}{\sqrt{2}} \left[V_{ub} V_{us}^*
\left(c_1 O_1^u + c_2 O_2^u \right) + V_{cb} V_{cs}^* \left(c_1
O_1^c + c_2 O_2^c \right) - V_{tb} V_{ts}^* \left( \sum_{i =
3}^{10} c_i O_i \right) \right] + {\rm H.c.}. \eea
 Here $c_i$'s are the wilson coefficients and the 4-quarks
current-current, gluonic and electroweak penguin operators are
defined by
 \begin{itemize} 
\item {\bf current-current operators}:
 \bea \label{eqn1:ccH}
O_1^u &=&
({\overline{u}} b)_{V-A} ({\overline {s}} u)_{V-A}
 ~~ ~~ O_2^u = ({\overline{u}_{\al}} b_{\bt})_{V-A} ({\overline {s}}_\bt
u_\al)_{V-A}, \nonumber\\
O_1^c &=& ({\overline{c}} b)_{V-A} ({\overline {s}} c)_{V-A}
 ~~ ~~ O_2^c = ({\overline{c}_{\al}} b_{\bt})_{V-A} ({\overline {s}}_\bt
c_\al)_{V-A},
 \eea
 \item {\bf QCD-penguin operators}:
  \bea \label{eqn2:ppGH}
O_3 &=& ({\overline{s}} b)_{V-A}
\sum_q({\overline {q}} q)_{V-A}, ~~ ~~ O_4 = ({\overline{s}_{\al}}
b_{\bt})_{V-A} \sum_q({\overline {q}}_\bt q_\al)_{V-A},
 \nonumber\\
  O_5 &=& ({\overline{s}} b)_{V-A}
\sum_q({\overline {q}} q)_{V+A}, ~~ ~~ O_6 = ({\overline{s}_{\al}}
b_{\bt})_{V-A} \sum_q({\overline {q}}_\bt q_\al)_{V+A}, \eea
 \item {\bf electroweak-penguin operators}:
 \bea \label{eqn3:ppEWH}
 O_7 &=& \frac{3}{2}({\overline{s}} b)_{V-A} \sum_q e_q({\overline {q}}
q)_{V+A}, ~~ ~~ O_8 = \frac{3}{2}({\overline{s}_{\al}}
b_{\bt})_{V-A} \sum_q e_q ({\overline {q}}_\bt q_\al)_{V+A},
 \nonumber\\
  O_9 &=& \frac{3}{2}({\overline{s}} b)_{V-A}
\sum_q e_q ({\overline {q}} q)_{V-A}, ~~ ~~ O_{10} =
\frac{3}{2}({\overline{s}_{\al}} b_{\bt})_{V-A} \sum_q e_q
({\overline {q}}_\bt q_\al)_{V-A},
\eea
\end{itemize}
where $\al, \bt$ are the $SU(3)$ color indices, $V\pm A$
correspond to $\gamma^\mu (1 \pm \gamma^5)$ and the wilson coefficients $c_i$'s, in which the QCD correction to weak interaction is encoded, are 
evaluated at the scale ${\cal O}(\mu\simeq m_b)$.
 $e$ and $g$
are respectively QED and QCD coupling constants and $T^a$'s are
$SU(3)$ color matrices. For the penguin operators, $O_3,
\dots,O_{10}$, the sum over $q$ runs over different quark flavors,
active at $\mu \simeq m_b$, i.e. $ q~ \epsilon \{u, d, s, c, b\}$. Note that 
the gluonic penguin operators $O_{3-6}$ contributes largely to $B\rightarrow \eta K$.
The contributions coming from the  electroweak penguin operators are not so significant and will be 
neglected in the present analysis.

 The wilson coefficients $c_i$s, quite well-known at the NLO order
accuracy, is evaluated at $\mu\simeq m_b$ and is within the 
perturbative control. The nontriviality (amounting uncertainity) exists in the evaluation 
of the hadronic matrix elements of the operators in the effective Hamiltonian 
comprising $B\rightarrow \eta K$ 
transition. Several approximations are made in order to have a control over 
this. 
Within naive factorization framework, one assumes the absence of the order 
${\cal O}(\al_s)$ QCD correction to the hadronic matrix element and the
 validity of 
working in the heavy quark limit (i.e. $m_b \gg \L_{QCD}$). 
Within the naive factorization, the hadronic matrix  of a $B$ meson decay to a pair of 
mesons $M_1$ and 
$M_2$ can be factorized and be written as the product of two quarks bilinear currents $j_1$ and $j_2$ 
\bea \label{eqn:naivef}
\langle M_1 M_2|O_i| {\overline B}\rangle_{NF}= \langle M_1|j_1| {\overline B}\rangle \times 
\langle M_2|j_2| 0\rangle
\eea 
which gives rise the transition form factor and the decay constant, respectively. 
The operators $O_i$, constituting the effective Hamiltonian, is specific to 
a particular $B$ decay. In the present case they are represented by 
Eqns.~(\ref{eqn1:ccH},\ref{eqn2:ppGH},\ref{eqn3:ppEWH}). Corrections to the 
naive factorization arises from the ``non-factorizable'' ${\cal O}(\al_s)$ QCD correction to the  hadronic matrix element by means 
of vertex, penguin, hard-spectator and as well as the weak 
annihilation contributions. The 
``QCD factorization'' framework, which can be looked as the order  ${\cal O}(\al_s)$ corrected version of the naive factorization framework, 
encodes all these
${\cal O}(\al_s)$ corrections to the hadronic matrix element 
(\ref{eqn:naivef}). Within QCD factorization, the above two body nonleptonic
decay amplitude (Eqn.~(\ref{eqn:naivef}), in the heavy quark limit, 
can be generalized as \cite{qcdf:BBNS}
\bea\label{eqn:qcdf2}
\langle M_1 M_2|O_i|\bar{B}\rangle_{QCDF} &=&
\sum_j F_j^{B\to M_1}(m_2^2)\,\int_0^1 du\,T_{ij}^I(u)\,\Phi_{M_2}(u)
\,\,+\,\,(M_1\leftrightarrow M_2)\nonumber\\
&&\hspace*{-2cm}
+\,\int_0^1 d\xi du dv \,T_i^{II}(\xi,u,v)\,
\Phi_B(\xi)\,\Phi_{M_1}(v)\,\Phi_{M_2}(u) 
\eea
where $M_1$ and $M_2$ corresponds to the light mesons. The factor 
$F^{B\rightarrow M_1 (M_2)}$ corresponds to the $B\rightarrow M_1$ transition form factor, while the 
$\Phi_X$ corresponds to the Light Cone Distribution Amplitudes (LCDA) 
for the meson 
$X$ for the quark-antiquark Fock states.
The perturbatively calculable 
quantities $T^{I,II}_{ij}$ in Eqn.~(\ref{eqn:qcdf2}), which are known as 
the hard-scattering kernels, contain essentially the ${\mathcal O}(\al_s)$ QCD
correction to the matrix element. $T^{I}_{ij}$ which starts at ${\cal O}(\al_s^{0})$, corresponds to the vertex, penguin correction, while $T^{II}_{ij}$ which starts at ${\cal O}(\al_s^{1})$ corresponds to the hard-spectator correction which arises due to the exchange of a hard gluon between the spectator quark inside the ${\overline B}$ meson with the quark of the emitted $M_2$ meson. Note that the weak annihilation contribution are not shown here, we will present them later seperately. Now no QCD correction implies the vanishing of 
the second term of Eqn.~(\ref{eqn:qcdf2}), while $T^{I}_{ij}$ turns into 
a constant. After performing the relevant integration 
(Eqn.~(\ref{eqn:qcdf2})) with the leading twist-2 LCDA's, one in the heavy 
quark limit ends up a expression, which much looks like: 
{\bf form factor $\times$ decay constnt}, agrees with the naive factorization 
result (Eqn.~(\ref{eqn:naivef})). Following the above discussion, the generic
${\overline B}\rightarrow M_1 M_2$ decay amplitude, 
in the heavy quark limt $m_b \gg \Lambda_{QCD}$ within the QCDF framwork, 
can be written as
\bea
\langle M_1 M_2|O_i|\bar{B}\rangle_{QCDF} = \langle M_1 M_2|O_i|\bar{B}\rangle_{NF} 
\left[1 + \sum_n r_n \al_s^n + {\cal O} \left(\frac{\Lambda_{QCD}}{m_b}\right)\right]. 
\eea
Note that in the heavy quark limt 
($m_b \gg \Lambda_{QCD}$) and at order ${\cal O}(\al_s)$, although the 
naive factorization framework breaks down (due to the presence of the second term), one can still calculates the  
the corrections systemetically by finding the corresponding corrections in the short-distance coefficients and the LCDA's of the mesons.

 ~In the QCDF framework, the amplitude for the ${\overline B}\rightarrow \eta K$ 
 can be expressed as
\bea \label{eqn:totamp}
{\cal M}_{SM}(\bar{B} \rightarrow \eta K) = {\cal M}^f_{SM}(\bar{B} \rightarrow \eta K) + {\cal M}^a_{SM}(\bar{B} \rightarrow \eta K),
\eea
where
\bea \label{eqn:totamp1}
{\cal M}^f_{SM}(\bar{B} \rightarrow \eta K) &=& \frac{G_F}{\sqrt{2}} \sum_{p=u,c}\sum_{i=1}^{i=6} \l_{p}~a^p_i~ 
\langle  \eta K|O_i| \bar{B} \rangle_{NF},
\eea
\bea \label{eqn:totamp2}
{\cal M}^a_{SM}(\bar{B} \rightarrow \eta  K)&=& \frac{G_F}{\sqrt{2}} f_B f_K f_{\eta } \sum_{p=u,c}\sum_{i=1}^{i=6} \l_{p}~b_i. 
\eea
In above 
$\l_p= V_{pb} V^*_{ps}$ with $V$'s being the CKM matrix elements, $G_F$, 
the Fermi decay constant, while $f_r$ ($r=B, K, \eta, \eta^{\prime} $) stands for the meson decay constants.
The non-factorizable vertex, penguin and hard spectator corrections are 
encoded in the effective coefficients $a^p_i$ which appeared in the 
first term of the amplitude (
Eqn.~(\ref{eqn:totamp})). The effective coefficients $a^p_i$ at the
 next-to-leading order in $\al_s(\mu)$ 
 takes the following form (for a generic nonleptonic 
$\bar{B}\rightarrow M_1 M_2$ decay) \cite{qcdf1:BN}
\bea \label{eqn:effcoef}
a^p_i (M_1,M_2)= \left(c_i(\mu) + \frac{c_{i\pm1}(\mu)}{N_c}\right) + \frac{c_{i\pm1}(\mu)}{N_c} 
\frac{C_F \al_s(\mu)}{4 \pi} \left[V_i(M_2) + \frac{4 \pi^2}{N_c} H_i(M_1, M_2)
\right] + P^p_i(M_2),
\eea 
where $c_i (\mu)$, the wilson coefficients with $i (=1,...6)$ is odd(even), the upper(lower) sign is applied and for the current-current operator the superscript $p$ is to be dropped (to avoid confusion). Here 
$N_c=3$ is the colour factor, $C_F=\frac{N_c^2 - 1}{2 N_c}$, the quadratic 
Casimir Invariant for $SU(N)$ ($N=N_c=3$ here) and $c_i$'s are the wilson 
coefficients. The quantities $V_i$, $H_i$ and $P^p_i$ respectively stands for the ``non-factorizable'' vertex, hard spectator and penguin corrections and their complete expressions can be found in  \cite{qcdf:BBNS},\cite{qcdf1:BN}. The hard spectator functions reads as
\begin{equation}\label{hardspecterms1}
   H_i(K \eta)
   = \frac{f_B f_K f^u_\eta}{m^2_B F_0^{B\rightarrow K}(0) 
f^u_\eta }\, \int_0^1 \frac{d\xi}{\xi} \Phi_B(\xi)\,
   \int_0^1\!dx \int_0^1\!dy \left[
   \frac{\Phi_{\eta}(x)\Phi_{K}(y)}{\bar x\bar y}
   + r_\chi^{K}\,\frac{\Phi_{\eta}(x)\Phi^{(3)}_{K}(y)}{x\bar y} \right]
\end{equation}
for $i=1$--4,
\begin{equation}\label{hardspecterms2}
   H_i(K \eta)
  = \frac{- f_B f_K f^u_\eta}{m^2_B F_0^{B\rightarrow K}(0) 
f^u_\eta }\, \int_0^1 \frac{d\xi}{\xi} \Phi_B(\xi)\,
   \int_0^1\!dx \int_0^1\!dy \left[
   \frac{\Phi_{\eta}(x)\Phi_{K}(y)}{x\bar y}
   + \,\frac{\Phi_{\eta}(x)\Phi^{(3)}_{K}(y)}{\bar x\bar y} 
\right]
\end{equation}
for $i=5$ with and $H_i(M_1M_2)=0$ for $i=6$. In above ${\bar j} = 1-j $ ~($j=x,y$) and the term 
$r_\chi^{K}=\frac{2 m_K^2}{m_b(\mu) (m_q + m_s)(\mu)}$, (where $m_q$ corresponds to the average of the up and down quark masses) is the chiral enhancement 
factor. The $\Phi_X (\Phi^{(3)}_X)$ corresponds to twist-2 (twist-3) LCDA's of the meson $X$. The weak annihilation contribution's $b_i$'s 
corresponding to Eqn.~(\ref{eqn:totamp2}), 
can be expressed as (following \cite{qcdf:BBNS},\cite{qcdf1:BN})
\bea \label{eqn:b1234}
   b_1 &=& \frac{C_F}{N_c^2}\,c_1 A_1^i \,, ~~b_3^p = \frac{C_F}{N_c^2} \left[ c_3 A_1^i + c_5 (A_3^i+A_3^f)
    + N_c c_6 A_3^f \right] \,, \\
 b_2 &=& \frac{C_F}{N_c^2}\,c_2 A_1^i \,,
    ~~b_4^p = \frac{C_F}{N_c^2}\,\left[ c_4 A_1^i + c_6 A_2^i \right] \,.
\eea
where $b_1, b_2$ corresponds to the current-current annihilation, while 
$b_3^p,~b_4^p$, the gluon penguin annihilation. Here $A_3^f$ is the factorizable annihilation amplitude which is being induced from the $(S-P)\times (S+P)$ type operators, whereas the non-factorizable annihilation amplitudes 
$A^{i}_{1,2,3}$ are induced from the $(V-A)\times (V-A)$, $(V+A)\times (V+A)$ and $(S-P)\times (S+P)$ type operators. Their explicit expressions, which
 can be found 
in  \cite{qcdf:BBNS},\cite{qcdf1:BN}, reads as
\begin{eqnarray}\label{blocks}
   A_1^i &=& \pi\alpha_s(\mu) \int_0^1\! dx dy\,
    \left\{ \Phi_{M_2}(x)\,\Phi_{M_1}(y)
    \left[ \frac{1}{y(1-x\bar y)} + \frac{1}{\bar x^2 y} \right]
    + r_\chi^{M_1} r_\chi^{M_2}\,\Phi^{(3)}_{M_2}(x)\,\Phi^{(3)}_{M_1}(y)\,
     \frac{2}{\bar x y} \right\} ,
    \nonumber\\
   A_1^f &=& 0 \,, \nonumber\\
   A_2^i &=& \pi\alpha_s(\mu) \int_0^1\! dx dy\,
    \left\{ \Phi_{M_2}(x)\,\Phi_{M_1}(y)
    \left[ \frac{1}{\bar x(1-x\bar y)} + \frac{1}{\bar x y^2} \right]
    + r_\chi^{M_1} r_\chi^{M_2}\,\Phi^{(3)}_{M_2}(x)\,\Phi^{(3)}_{M_1}(y)\,
     \frac{2}{\bar x y} \right\} ,
    \nonumber\\
   A_2^f &=& 0 \,, \\
   A_3^i &=& \pi\alpha_s(\mu) \int_0^1\! dx dy\,
    \left\{r_\chi^{M_1}\,\Phi_{M_2}(x)\,\Phi^{(3)}_{M_1}(y)\,
    \frac{2\bar y}{\bar x y(1-x\bar y)}
    - r_\chi^{M_2}\,\Phi_{M_1}(y)\,\Phi^{(3)}_{M_2}(x)\,
    \frac{2x}{\bar x y(1-x\bar y)} \right\} , \nonumber\\
   A_3^f &=& \pi\alpha_s(\mu) \int_0^1\! dx dy\,
    \left\{r_\chi^{M_1}\,\Phi_{M_2}(x)\,\Phi^{(3)}_{M_1}(y)\,
    \frac{2(1+\bar x)}{\bar x^2 y}
    +  r_\chi^{M_2}\,\Phi_{M_1}(y)\,\Phi^{(3)}_{M_2}(x)\,
    \frac{2(1+y)}{\bar x y^2} \right\}, \nonumber
\end{eqnarray}
for a generic $\bar{B}\rightarrow M_1 M_2$ decay. In our case of interest,
the corresponding expressions are obtained by setting $M_1=K (\eta)$ and 
$M_2=\eta (K) $.  
Note that the power suppressed twist-3 LCDA $\Phi^{(3)}_X$ of the hard spectator and annihilation 
contributions suffers from the end point divergence 
$X_{H,A} = \int_0^1 \frac{d y}{y}$, which can be phenomenolgically parametrized as \cite{qcdf1:BN}
\bea
X_H = \int_0^1 \frac{d y}{y} = (1 + \rho_{H,A} e^{i \phi_{H, A}})~ln\left(\frac{m_B}{\L_{h}}\right)
\eea
where $\Lambda_h \sim 0.5$ GeV, $\rho_{H,A}$ are free parameters 
expected to be about of order
$\rho_{H,A} \simeq {\cal O}(1)$ and $\phi_{H, A}~ \epsilon ~[0, 2\pi]$.
With all these QCD corrections in hand, we are now ready to 
give the ${\cal O}(\al_s)$ corrected expressions for the 
$\bar{B}\rightarrow \eta K $ decay 
amplitude (i.e. Eqn.~(\ref{eqn:totamp1})) and within QCDF they 
reads (for $\bar{B} = B^-$) as \cite{b2etap:Ali}
\bea \label{eqn:SMb2etaKF}
{\cal{M}}^f_{SM} (B^- \rightarrow \eta K^-) &=& \frac{G_F}{\sqrt{2}} \, \left\{ V_{ub} V_{us}^* \, \left[
a^p_2 + a^p_1 \frac{m_B^2-m_{\eta}^2}{m_B^2-m_K^2} \,
\frac{F_0^{B \to \eta}(m_K^2)}{F_0^{B \to K^-}(m_{\eta}^2)} \,
\frac{f_K}{f_{\eta}^u} \right]
+ V_{cb}V_{cs}^* \, a^p_2 \frac{f_{\eta}^{c}}{f_{\eta}^u}
\right. \nonumber \\
&&- V_{tb} V_{ts}^* \, \left[2 a^p_3 -2 a^p_5 + \left( a^p_3 -a^p_5 + a^p_4
+ \frac{ a^p_6 m_{\eta}^2}{ m_s(m_b-m_s)} \right) \,
\frac{f_{\eta}^s}{f_{\eta}^u}
- \frac{ a^p_6 m_{\eta}^2}{ m_s(m_b-m_s)}  \,
\right. \nonumber \\
&& \left. \left. +
\left( a^p_4 + \frac{2 a^p_6 m_K^2}{(m_s+m_u) \, (m_b-m_u)} \right)
\, \frac{m_B^2-m_{\eta}^2}{m_B^2-m_K^2} \, \frac{F^{B \to
\eta}_0(m_K^2)}{F^{B \to K^-}_0(m_{\eta}^2)} \,
\frac{f_K}{f_{\eta}^u} \right] \,
\right\} \, \bra K^- | \bar{s} \, b_-|B^- \ket \,
            \bra \eta | \bar{u} \, u_-|0 \ket \nonumber \\
\eea
where $b_-$ (and $u_-$ ) corresponds to $\g_\mu (1 - \g_5) b$ 
(and $\g_\mu (1 - \g_5) u$). 
The transition form factor and decay constants are defined as 
\bea
\bra K^-(p') | \bar{s} \, b_-|B^-(p) \ket &=& \bra K^-(p') | \bar{s} \, \g^\mu b|B^-(p) \ket \nonumber \\
&=& \left[\left\{(p + p')^\mu - \frac{m^2_B - m^2_K}{q^2} q^\mu \right\} F^{B\to K}_1 (q^2) +  \frac{m^2_B - m^2_K}{q^2} q^\mu F^{B\to K}_0 (q^2)\right]
\eea
\bea
\bra \eta(q) | \bar{u} \, u_-|0 \ket = -\bra \eta(q) | \bar{u} \, 
\g_\mu \g_5 u |0 \ket = i f^u_{\eta} q_\mu.
\eea
where $m_B,~m_K$ corresponds to $B,K$ meons masses and $f^u_\eta$, the $\eta$ meson decay
constant. $F^{B\to K}_{0,1} (q^2)$ are the $B \to K$ transition form factors evaluated
at $q^2 = m^2_\eta$. Note that in the effective coefficients $a^p_i$ 
(Eqn.~(\ref{eqn:effcoef})) appearing in the above amplitude (Eqn.~(\ref{eqn:SMb2etaKF}))  contains the non-factorizable vertex ($V_i$) and penguin ($P^p_i$) corrections, which  
are evaluated at $\mu = m_b~(m_b/2)$ and the hard spectator ($H_i$) and annihiation corrections (given by Eqn.~(\ref{eqn:b1234})), evaluated  at  
$\mu = \mu_h \sim \sqrt{\mu \L_h}$, where $\L_h \simeq 0.5$ GeV 
\cite{qcdf1:BN}.
It is now straightforward to write down the branching ratio (BR) for 
the 
$B^- \rightarrow \eta K^-$ decay as
\bea \label{eqn:bratio}
  BR (B^- \rightarrow \eta K^-) = \frac{\tau_B p_c}{8 \pi m_B^2} 
{|{{\cal {M}}(B^- \rightarrow \eta K^-)}|^2} 
\eea
where ${\cal {M}}(B^- \rightarrow \eta K^-)$ is the decay amplitude, $\tau_B$, the $B$ meson life time 
and $p_c$, the c.m. momentum of the $\eta$ and $K^-$ mesons in the $B^-$ rest frame, is given by
\bea
p_c = \frac{\sqrt{\left(m_B^2 - (m_{\eta} - m_{K^-})^2\right)
~\left(m_B^2 - (m_{\eta} + m_{K^-})^2\right)}}{2 m_B}.
\eea
The BR's $BR (B^+ \rightarrow \eta K^+)$ is obtained from 
$BR (B^- \rightarrow \eta K^-)$ by performing the relevant changes 
under CP transformations.
 The CP-asymmetry $A_{cp}$, according to the standard convention, 
is defined as 
\bea
A_{cp} = \frac{\Gamma(B^- \to \eta {K^-}) - \Gamma(B^+ \to\eta {K^+} )}
{\Gamma(B^- \to \eta {K^-}) + \Gamma(B^+ \to\eta {K^+} )}.
\eea

\section{Numerical Analysis}\label{sec:section3}

The BR and CP-assymmetry $A_{cp}$ data for the $B^{+} \to \eta ~K^+$ decay 
are available in the HFAG05 website \cite{HFAG05} and is given in 
Table {\ref{tab:table1}}.
\begin{table}[htb]
\caption{\label{tab:table1} Data for the $B^+ \to \eta K^+$ decay mode. 
The error bars are at $1 \sigma $ limit; the upper limit at $90\%$ CL.For $A_{cp}$ we follow the standard convention as mentioned in the text.}
\begin{center} \label{HFAG05}
\begin{tabular}{|l|c|c|c|}
\hline
Final State& $BR \times 10^{6}$& $A_{cp}$ \\
\hline
$\eta K^+$&$2.6 \pm 0.5$&$-0.25\pm 0.14$\\
\hline
\end{tabular}
\end{center}
\end{table}
The decay amplitudes depend on the effective coefficients $a_i$'s,
CKM matrix elements, quark masses and the non-perturbative inputs like 
hadronic form factors, decay constants etc.
\subsection{CKM matrix elements}
The parameters $A,
\l, \rho$ and $\eta$ of the CKM matrix $V_{CKM}$ (\ref{eqn:ckm}) in the 
Wolfenstein parametrization are set at the following values.
We employ $A$ and $\l = \sin\theta_c$ at the values of $0.815$
and $0.2205$ in our analysis. The other parameters
are found to be $ \rho = \sqrt{{\overline \rho}^2 + {\overline \eta}^2}
~cos{\gamma}$ and $ \eta = \sqrt{{\overline \rho}^2 +
{\overline \eta}^2}~ sin{\gamma}$ with $\gamma = arg(V^*_{ub})$ (an unknown 
 parameter in our analysis) and 
$\sqrt{{\overline \rho}^2 + {\overline \eta}^2}
= 0.3854$\cite{PDG}. Here ${\overline \rho} = \rho (1 - \frac{\lambda^2}{2})$ and
${\overline \eta} = \eta (1 - \frac{\lambda^2}{2})$ \cite{brev:bbl}.
\subsection{Effective coefficients $a_i$, quark masses, decay constants
 and form factors }
The NLO wilson coefficients in the NDR scheme, obtained in the paper by
Beneke {\etal} \cite{qcdf1:BN},are cataloged in Table~\ref{tab:effectiveci}.
\begin{table}[tbh]
\caption{$\Delta B = 1$ wilson coefficients at $\mu =\frac{m_b}{2} (m_b)
\sim 2.1 (4.2)$ GeV for $m_t=170$ GeV, $\al = 1/129$ and
$\Lambda^{(5)}_{\overline {MS}}=225$ MeV  in the NDR scheme~\cite{qcdf:BBNS}.}
\label{tab:effectiveci}
\begin{center}
\begin{tabular}{|c||c|c|c|c|c|c|c|c|c|c|}
\hline
\hline NLO& $c_1$ &$c_2$ &$c_3$ &$c_4$ &$c_5$ &$c_6$ \\
\hline
$\mu=m_b/2$ &$1.137$ &$-0.295$ &$0.021$&$-0.051$& $0.010$&$-0.065$ \\
\hline
$\mu=m_b$ &$1.081$ &$-0.190$ &$0.014$&$-0.036$& $0.009$&$-0.042$ \\
\hline
\end{tabular}
\end{center}
\end{table}
The effective coefficients $a_i$'s ($i=1,2,...,10$) in 
Eqn.(\ref{eqn:SMb2etaKF}), can be obtained by using Eqn.(\ref{eqn:effcoef})
from these NLO wilson coefficients.
The wilson coeffficients $c_i(\mu)$ and coupling constant 
$\al_s(\mu)$ (in Eqn.~(\ref{eqn:effcoef})) 
are evaluated at $\mu = m_b (m_b/2)$. The scale chosen for the vertex, 
penguin, hard-spectator and annihilation term are described above (see 
the discussion before Eqn.~(\ref{eqn:bratio})). 

~For the quark's constituent masses, we use $m_b=4.2$ GeV, 
$m_c=1.5$ GeV and 
$m_s=0.50$ GeV and $m_u = m_d \sim 0.2$ GeV, the scale independent quantities
\cite{qcdf1:BN,b2etap:Ali}, which appear in the loop integral and for
the current masses, we use their scale dependent values as listed in 
Table \ref{tab:currentmass}. 
\begin{table}[htbp]
\caption{ The scale dependent current quark masses which are taken from \cite{qcdf1:BN},\cite{b2etap:Ali}.}
\label{tab:currentmass}
\begin{center}
\begin{tabular}{|c|c|c|c|c|}
\hline
\hline & $m_b(\mu)$ &$m_s(\mu)$ &$m_u(\mu)$ &$m_d(\mu)$ \\
\hline
$\mu = m_b/2$& $4.88$& $0.122$& $0.0413\times m_s(\mu)$ & $0.0413\times m_s(\mu)$\\
\hline
$\mu = m_b$& $4.20$& $0.090$& $0.0413\times m_s(\mu)$ & $0.0413\times m_s(\mu)$\\
\hline
\end{tabular}
\end{center}
\end{table}
which appear in the factorized amplitude after making the use of equation of motion of quarks.

~The {\it decay constants} (in GeV) are given by (\cite{qcdf1:BN},\cite{b2etap:Ali})
\bea
f_B=0.20,~f_\pi=0.131,~f_K=0.160,~f^c_{\eta'}=0.0058,~f^c_\eta=0.00093
\eea
and we use $f_0 = 1.10 f_\pi$ and $f_8=1.34 f_\pi$ with the one angle mixing scheme
(in $\eta-\eta'$ sector) to obtain
\bea
f^u_\eta = \frac{f_8~cos\theta}{\sqrt{6}} - \frac{f_0~sin\theta}{\sqrt{3}},~~f^s_\eta = -2 \frac{f_8~cos\theta}{\sqrt{6}} - \frac{f_0~sin\theta}{\sqrt{3}}
\eea
 with $\theta$, the mixing angle, another unknown parameter in our 
analysis.

~We use the following non-perturbative {\it form factors} values in our 
analysis \cite{qcdf1:BN}, 
\bea
F_{0,1}^{B\to \pi}(0) = 0.28,~F_{0,1}^{B\to K}(0) = 0.34,\nonumber \\
~F_{0,1}^{B\to \eta}(0) = F_0^{B\to \pi}(0) \left[\frac{cos\theta}{\sqrt{6}} - \frac{sin\theta}{\sqrt{3}} \right] .
\eea
For the distribution amplitudes (LCDA), we use the asymptotic form for the pseudoscalar 
mesons \cite{qcdf1:BN} 
\bea
\Phi_\eta &=& 6 x (1 - x), ~ \Phi_K = 6 x (1 - x), ~twist-2 ~LCDA \\
\Phi^{(3)}_\eta &=& 1, ~ \hspace*{0.45in} \Phi^{(3)}_K = 1, \hspace*{0.45in} ~twist-3 ~LCDA. 
\eea
The $B$ meson wave function 
\bea
\Phi_B (\bar{\rho}) = N_B {\bar{\rho}}^2 (1 - \bar{\rho})^2 exp\left[-\frac{1}{2} \left(\frac{\bar{\rho} m_B}{\omega_B}\right)^2\right],
\eea
with the normalization constant $N_B$ is being determined from the condition
\bea
\int_0^1 \Phi_B (\bar{\rho}) d \bar{\rho} = 1, 
\eea
with $m_B= 5.278$ GeV and $\omega_B = 0.25$ GeV \cite{bwave:chengyang}.

\subsection{Results and Discussions}
With all the numerical inputs in hand, we are now ready to discuss our results. Working within the QCD factorization framework, we made use of 
the $BR(B^+\to \eta K^+)$ and CP-asymmetry $A_{cp}(B^+\to \eta K^+)$ 
\cite{HFAG05} 
data to obtain constraints in the 
$\theta~-~\gamma$ plane. As we will see that the 
bound depends crucially on the scale 
$\mu$ at which different 
nonfactorizable ${\cal{O}} (\al_s)$ corrections are evaluated. The 
impact of the nonfactorizable 
{\it weak annihilation} correction, which is usually power suppressed,  
will also be considered while obtaining such constraints.   

\subsubsection{Constraints in the $\theta - \gamma$ plane: with (without) 
the annihilation terms} 
We have seen that the     
effective coefficients $a_i$'s (Eqn.\ref{eqn:effcoef}), 
contains the wilson coefficients $c_i$'s, coupling constant $\al_s (\mu)$ and 
 several nonfactorizable corrections,
which crucially depends on the renormalization 
scale $\mu$.  
For the wilson coefficients $c_i$ in $a_i$, we use their 
NLO values for different $\mu$ which are displayed 
in Table \ref{tab:effectiveci}.  The current quark 
mass which depends on the scale $\mu$, are displayed in 
Table \ref{tab:currentmass}. For the running coupling $\al_s(\mu)$, we use 
the oneloop expression for $\al_s (\mu)$ as  \cite{brev:bbl}
\bea
\al_s (\mu) = \frac{4 \pi}{\bt_0 Log \left[\mu^2/\Lambda_{\overline{MS}}^2 \right]},
\eea
where $\beta_0 = \left(\frac{11}{3} N_c - \frac{2}{3} n_f \right)$ with $N_c = 3$, the colour and the typical QCD scale parameter 
$\Lambda_{\overline{MS}} (\Lambda_{QCD}) = 226$ MeV with $n_f = 5$, the number 
of active quark flavour at $\mu= m_b$ and similarly $\Lambda_{\overline{MS}} (\Lambda_{QCD}) = 372$ MeV with $n_f = 4$ for $\mu= m_b/2$ \cite{brev:bbl}.
Using these as inputs, we now use the BR and $A_{cp}$ data ( Table 
\ref{HFAG05}) 
to obtain constraint in the $\theta$ and $\gamma$ plane. Before
to obtain such constraints, let us see how the CP asymmetry 
$A_{cp}$ varies with $\gamma$ or $\theta$. 
In Figures 1(a,b) and 2(c,d), we made such plots. From Figures 1(a,b) ( 
standing respectively for $\mu = m_b/2$ and $m_b$ with the mixing 
angle $\theta = -21.3^o$ \cite{b2etap:Ali}), we see that within the 
$\pm 1\sigma$ deviation
with respect to the the central value $A^{cen}_{cp}= -0.25$, 
$\gamma$ is allowed to varry in between $+20^o$ and $+100^o$ (one range), 
$140^o$ and $+170^o$ 
(other range). The negative $\gamma$ is ruled out at the $\pm 1\sigma$ level as far  
as the CP asymmetry data is concerned. From  Figures 2(c,d), in which the unitarity angle $\gamma$ is chosen as $+70^o$ \cite{qcdf1:BN}, the scale $\mu$, is being set at  
$m_b/2$ and $m_b$, respectively, we find that at the $\pm 1\sigma$ error away from the central value $A^{cen}_{cp}$, the mixing angle $\theta$ is allowed to lie in between $-45^o$ and $-40^o$ (one range), $-25^o$ and $-10^o$ 
(other range). However, the $A_{cp}$ data allows another region, in which 
$\theta$ is positive. However, the constraints on $\theta$, which follows from 
the $\eta \to \gamma \gamma$, $\eta' \to \gamma \gamma$ and $\pi^0 \to \gamma \gamma$ decay studies, suggest that $\theta$ should be negaive i.e.
\bea
\theta = (-21.3 \pm 2.5)^o.
\eea
In order to have the overall consistency with other's analysis,  
we will not explore the {\it positive} $\theta$ region. From the above analysis, where
one angle is varied, while the other fixed, we know their 
sizes and signs.  We will now explore  
the general possibility in which both $\theta$ and $\gamma$ is allowed to vary.
 In this case, by using the $A_{cp}(B^+ \to \eta K^+)$ data, as well as the 
$BR(B^+ \to \eta K^+)$ data, it is possible to constrain the two angles $\theta$ and 
$\gamma$ simultaneously by finding the contour plots in the $\theta-\gamma$ plane.

In Figures 3(e,~f), we obtain such contour plots corresponding to $\mu = m_b/2$ 
and $m_b$.  
The central curve corresponds to the region allowed by the 
$A_{cp} = -0.25$ (i.e. the central value), while the next(outer) to the 
central curve corresponds to $A_{cp} = -0.11(=-0.25 +0.14)$ and the region between the outer and central curves, is allowed at the $+1\sigma$ level. 
The innermost curve 
corresponds to $A_{cp} = -0.39$ and the region between this curve and the central one, is allowed at the $-1\sigma$ level. Note that a different set of $\mu$
 choice results into the change in the shape and hence the area of the allowed regions as a whole. At the $\pm 1\sigma$ level, in either case, quite a large range of both $\theta$ and $\gamma$'s are allowed: e.g. while the mixing angle
$\theta$ varies in between $-42^o$ to $-6^o$, the unitarity angle $\gamma$ can changes largely in between, say, e.g. $10^o$ to $175^o$ at the $\pm 1\sigma$ level. As for example, say $\gamma = 70^o$, the contour plot 
suggest that  $\theta$ should lie in between $-44^o$ and $-40^o$ (left zone), $-24^o$ and $-6^o$ (right zone) for the scale 
$\mu=m_b/2$, while in between$-42^o$ and $-40^o$(left zone), $-26^o$ and $-8^o$ 
(right zone) for the scale $\mu=m_b$.

In Figures 4(g,~h), the $\theta~-~\gamma$ contour plots corresponds to 
$BR(B^+ \to \eta K^+) = (2.6\pm 0.5) \times 10^{-6}$. The 
renormalization scale $\mu$ is chosen at $m_b/2$ (Figure 4g) and $m_b$ (Figure 4h). There are two regions in this plot. For each region, 
the central curve corresponds to the central value 
$BR(B^+ \to \eta K^+) = 2.6 \times 10^{-6}$, while the next(outer i.e. 
rightwards for the right side, and leftwards in the left side) to the 
central curve corresponds to $BR(B^+ \to \eta K^+) = 3.1 \times 10^{-6}$ and the region between the outer and central curves (on either side), 
is allowed at the $+1\sigma$ level. 
The two innermost curves
corresponds to $BR(B^+ \to \eta K^+) = 2.1 \times 10^{-6}$ and the region between this curve and the central one (in either side), 
is allowed at the $-1\sigma$ level. The allowed contour region followed from 
the BR data, is rather narrow than that obtained from the $A_{cp}$ data. 
Within $\pm 1\sigma$ error bar, while $\gamma$ can change widely in between 
$0^o$ to $180^o$, $\theta$ changes very little: it change only by $\pm 1^o$ with respect to the central value (corresponding to the central curve).
Note that a different $\mu$ shift the $\theta~-~\gamma$ allowed contour region 
(see Figure 4g and 4h).
As an example, say $\gamma = 70^o$, the contour plot suggest that  $\theta$ should lie in between $-46^o$ and $-44^o$ (left region) and in between 
 $-30^o$ and $-26^o$ (right region) for the scale 
$\mu=m_b/2$, while for $\mu=m_b$, it lies in between 
$-54^o$ and $-50^o$ (left region) and  $-30^o$ and $-26^o$ (right region).

To see the explicit scale dependence of the BR and CP asymmetry $A{cp}$ 
results, 
let us calculate them for some chosen $\mu$ values. 
At $\mu = m_b/2$, with $\theta = -21.3^o$ and $\gamma = 70^o$, a point well 
within the allowed region, we find 
$BR(B^{-(+)} \to \eta K^{-(+)}) = 2.98(5.72) \times 10^{-6}$ and 
$A_{cp}(B^- \to \eta K^-) = -0.315$ when the annihilation
term is taken into consideration. If the annihilation term is switched  off, the correponding predictions turns out to be 
$BR(B^{-(+)} \to \eta K^{-(+)}) = 3.0(5.83) \times 10^{-6}$ 
(a change about $2\% (11\%)$) and 
$A_{cp}(B^- \to \eta K^-) = -0.319$ (a change of $\sim 4\%$).
Similarly, for $\mu = m_b$, with the same $\gamma$ and $\theta$ value and 
 considering the the annihilation contribution, 
we find $BR(B^{-(+)} \to \eta K^{-(+)}) = 
2.93(4.91) \times 10^{-6}$ and 
$A_{cp}(B^- \to \eta K^-) = -0.252$. Without the annihilation term, we find 
$BR(B^{-(+)} \to \eta K^{-(+)}) = 
2.94(4.93) \times 10^{-6}$ (a change about $1\% (2\%)$) and 
$A_{cp}(B^- \to \eta K^-) = -0.253$ (a change of $\sim 1\%$). 
So, the impact of power correction (annihilation contribution) 
on $BR$ and $A_{cp}$ at different $\mu$ is not so significant and consequently, 
in constraining  the parameter space of $\theta$ and $\gamma$. 

In Figures 5(i,~j) and 6(k,~l), we have plotted 
$A_{cp}$ and $BR(B^+ \to \eta K^+)$ as a function of $\theta$ and $\gamma$ 
with different set of the renormalization scale i.e.
$\mu = m_b/2$ (left one) and $m_b$ (right one). Note the differences arises due to the choice in renormalization scale: in Figures 5(i,j) larger negative 
$A_{cp}$ may arise at $\mu = m_b$ than that due to $\mu = m_b/2$ and in 
Figures 6(k,l), in the same way large BR also arise at $\mu = m_b/2$ in comparison to $\mu = m_b$. Finally, from the Figures  5(i,~j), we see that 
a large negative $A_{cp}$, consistent with the present data, allows $\theta$ to lie 
around $-40^o$, for a wide range 
of $\gamma$, consistent with our finding. 

\section{Summary and Conclusion} \label{sec:section4}
We have investgated the charmless nonleptonic $B\to \eta K$ decay within the QCD factorization framework in the Standard Model. Two crucial things in the decay amplitudes are: (i) the unitarity angle $\gamma$, and (ii) the $\eta-\eta'$ mixing angle $\theta$, which 
  possesses several independent constraints followed from different studies. 
By knowing either $\theta$ or $\gamma$ from others analysis and using the HFAG $A_{cp}$ data, one can get some idea about the allowed range of $\gamma$ or $\theta$. 
We considered the most general scenario in which both $\theta$ and $\gamma$ are 
being treated as the unknown 
parameters. We use the $BR(B^+ \to \eta K^+)$ and $A_{cp}(B^+ \to \eta 
K^+)$ (available in the HFAG webcite) 
to constrain the $\theta - \gamma$ contour plane. 
Using the $A_{cp}$ data, at the 
$\pm 1 \sigma$ level, we found that when the angle $\theta$ varies in between $-42^o$ 
to $-6^o$ (in two ranges), the angle $\gamma$ can vary in between $0^o$ to $180^o$. For the $BR(B^+ \to \eta K^+) = (2.6 \pm 0.5)\times 10^{-6}$, 
with $\gamma = 70^o$, we found that 
$\theta$ can vary in between $-46^o$ and $-44^o$, $-30^o$ and $-26^o$ for 
$\mu = m_b/2$. For $\mu = m_b$, the corresponding allowed range is in between
$-54^o$ and $-50^o$, $-30^o$ and $-26^o$ with the same choice of $\gamma$. 
 We find the diference in the allowed contour region of  
$\theta$ which follows from the $A_{cp}$ and $BR$ constraints, 
a crucial finding of the present study.
We made our analysis for different set of 
renormalization scales i.e. $\mu$'s, which results into change in the 
allowed region in the $\theta - \gamma$ contour plane. This is an important 
observation of this work. 
The effect of power correction correction 
is also investigated and found to be not so significant in constraining the 
$\theta -\gamma$ contour plane. This is so because it's (power correction) 
impact on $A_{cp}$ and 
$BR(B^{-(+)} \to \eta K^{-(+)})$ are about $1\%$ and 
$1\% (2\%)$, respectively.\\

{ {\large {\it Acknowledgments}}}~
I would like to thank the physics department of CYCU, Chung-Li, 
Taiwan (R.O.C) for the hospitality where a part of this work is 
being completed. 


\vspace*{-0.5in}
\newpage
\begin{figure}
\subfigure[]{
\label{PictureOneLabel}
\hspace*{-0.7 in}
\begin{minipage}[b]{0.5\textwidth}
\centering
\includegraphics[width=\textwidth]{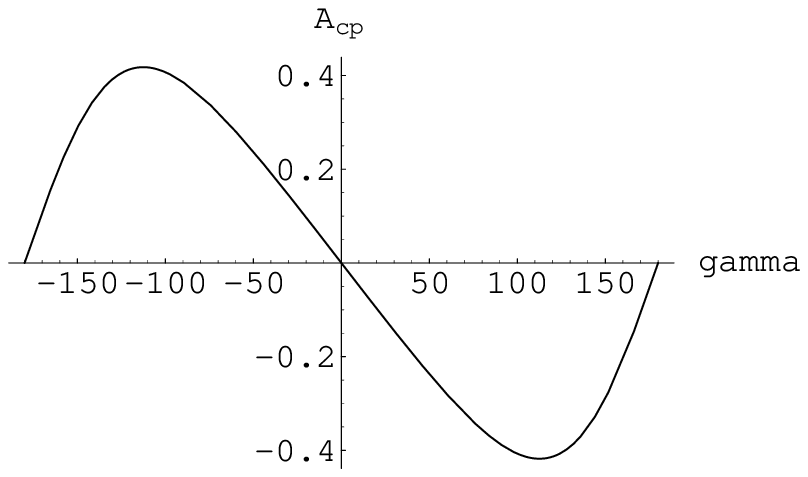}
\end{minipage}}
\subfigure[]{
\label{PictureTwoLabel}
\hspace*{0.3in}
\begin{minipage}[b]{0.5\textwidth}
\centering
\includegraphics[width=\textwidth]{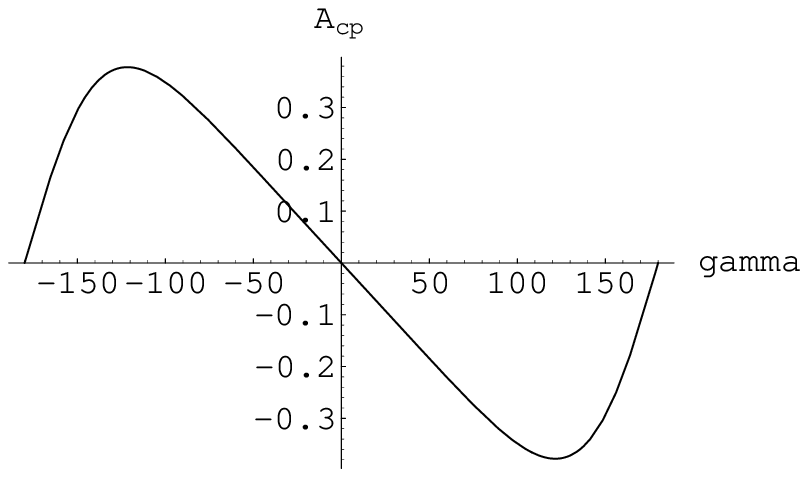}
\end{minipage}}
\end{figure}
\vspace*{-0.45in}
\noindent {\bf Figure 1[a,~b]}.
{{ \it Plot showing the CP asymmetry $A_{cp}(B^+ \to \eta K^+)$ as a function of 
the unitarity angle gamma ($\gamma$, given in degree). 
The most recent data \cite{HFAG05} for 
$A_{cp} = -0.25 \pm 0.14$. The Figure 1a and 1b corresponds to the 
following choice of the renormalization scale: $\mu=\frac{m_b}{2}$ (Fig.1a) and $\mu=m_b$ (Fig.1b).}}
\vspace*{-0.5in}
\newpage
\begin{figure}
\subfigure[]{
\label{PictureOneLabel}
\hspace*{-0.7 in}
\begin{minipage}[b]{0.5\textwidth}
\centering
\includegraphics[width=\textwidth]{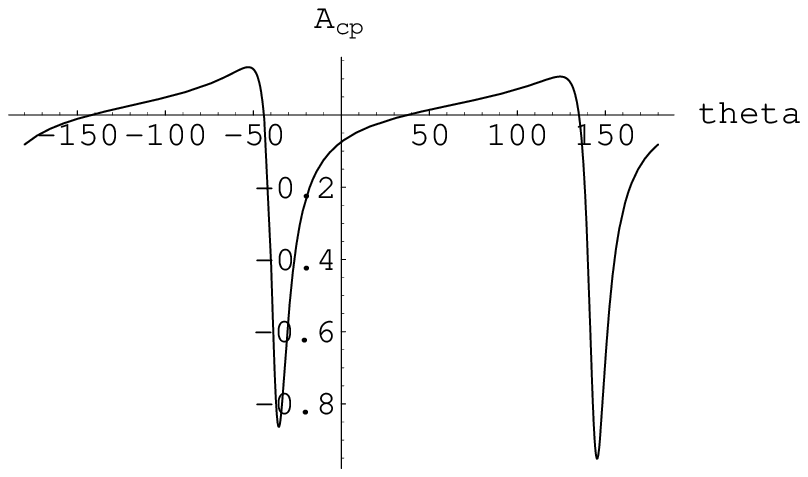}
\end{minipage}}
\subfigure[]{
\label{PictureTwoLabel}
\hspace*{0.3in}
\begin{minipage}[b]{0.5\textwidth}
\centering
\includegraphics[width=\textwidth]{AcpTheta2.eps}
\end{minipage}}
\end{figure}
\vspace*{-0.45in}
\noindent {\bf Figure 2[c,~d]}.
{{ \it Plot showing the CP asymmetry $A_{cp}(B^+ \to \eta K^+)$ as a function of 
the $\eta$-$\eta'$ mixing angle theta ($\theta$, given in degree). 
The CP-asymmetry $A_{cp} = -0.25 \pm 0.14$ accoriding to the HFAG webcite 
\cite{HFAG05}. The Figure 2c and 2d corresponds to the choice of the renormalization scale: $\mu=\frac{m_b}{2}$ (Fig.2c) and $\mu=m_b$ (Fig.2d).}}
\vspace*{-0.5in}
\newpage
\begin{figure}
\subfigure[]{
\label{PictureOneLabel}
\hspace*{-0.7 in}
\begin{minipage}[b]{0.5\textwidth}
\centering
\includegraphics[width=\textwidth]{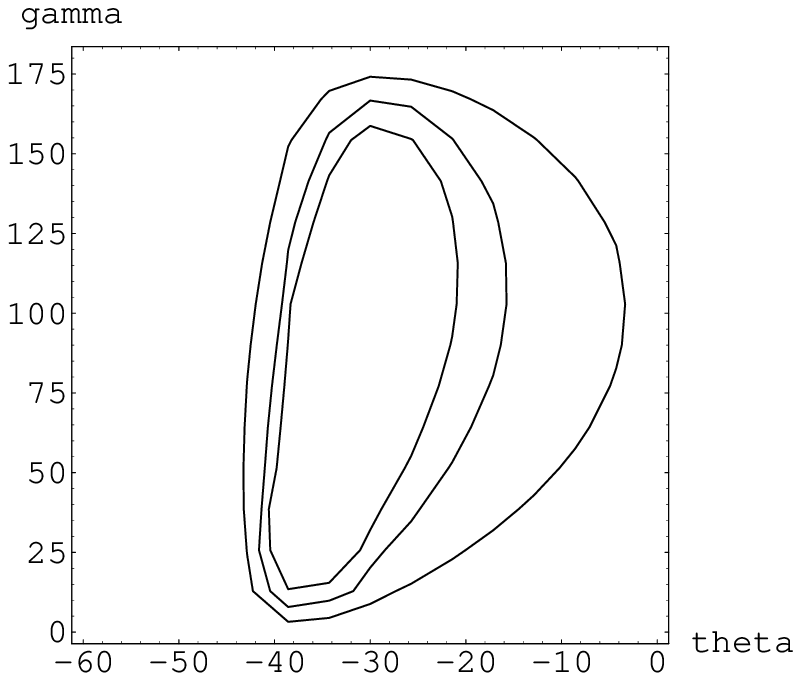}
\end{minipage}}
\subfigure[]{
\label{PictureTwoLabel}
\hspace*{0.3in}
\begin{minipage}[b]{0.5\textwidth}
\centering
\includegraphics[width=\textwidth]{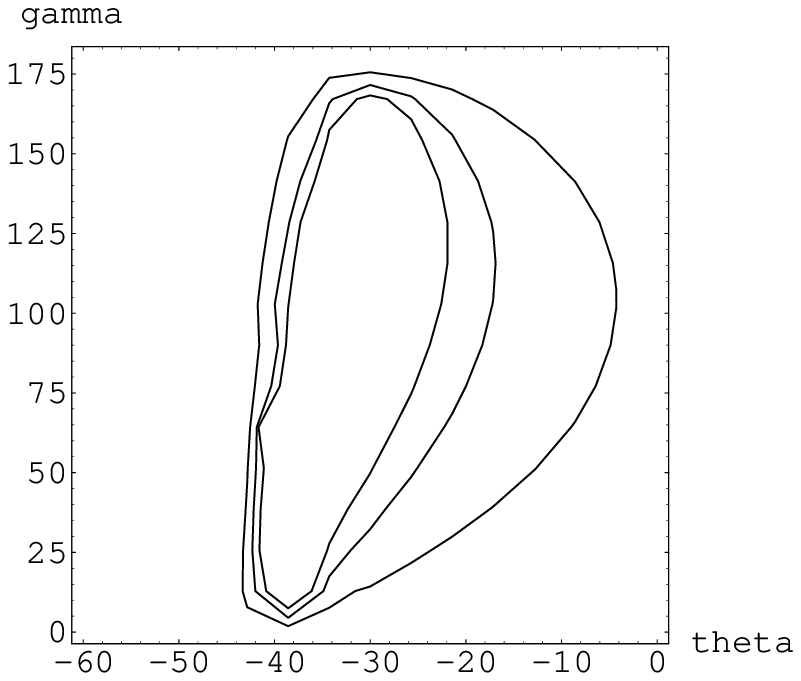}
\end{minipage}}
\end{figure}
\vspace*{-0.45in}
\noindent {\bf Figure 3[e,~f]}.
{{\it Contour plots in the theta ($\theta$) - gamma ($\gamma$) plane (angles are given 
in degree) using $A_{cp} = -0.25 \pm 0.14$ \cite{HFAG05} with the renormalization scale 
$\mu=m_b/2$ (Fig.3e) and $m_b$ (Fig.3f), respectively.}}
\vspace*{-0.5in}
\newpage
\begin{figure}
\subfigure[]{
\label{PictureThreeLabel}
\hspace*{-0.7 in}
\begin{minipage}[b]{0.5\textwidth}
\centering
\includegraphics[width=\textwidth]{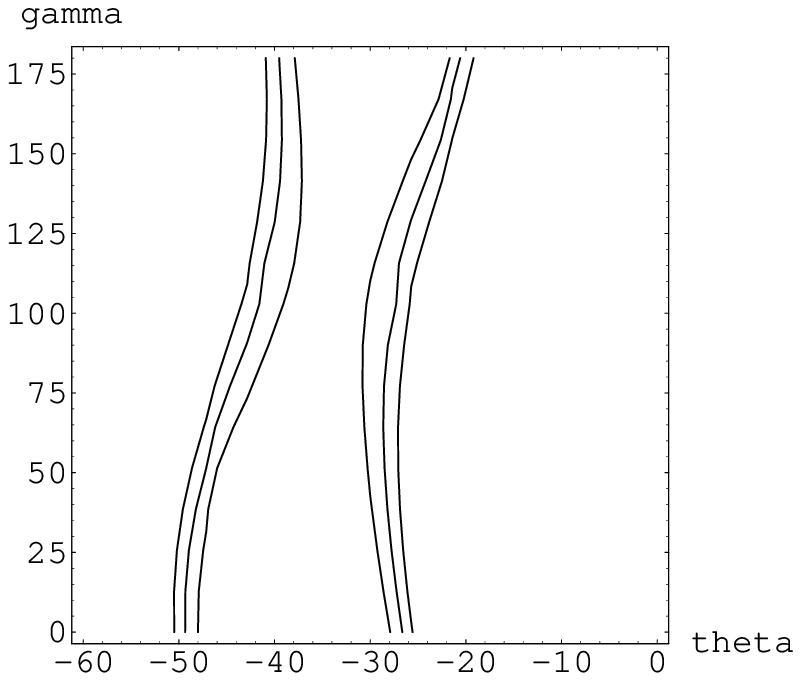}
\end{minipage}}
\subfigure[]{
\label{PictureFourLabel}
\hspace*{0.3in}
\begin{minipage}[b]{0.5\textwidth}
\centering
\includegraphics[width=\textwidth]{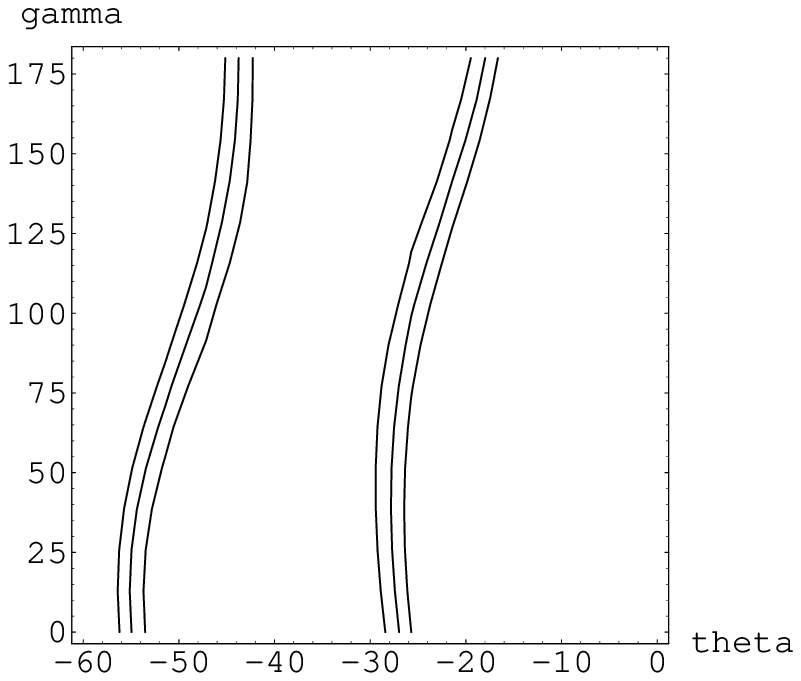}
\end{minipage}}
\end{figure}
\vspace*{-0.45in}
\noindent {\bf Figure 4[g,~h]}.
{{ \it  Contour plots in the theta ($\theta$) - gamma ($\gamma$) plane  
(angles are given in degree) using 
$BR(B^+ \to \eta K^+) = (2.6 \pm 0.5)\times 10^{-6}$ \cite{HFAG05} with the renormalization scale $\mu=m_b/2$ (Fig.4g) and $m_b$(Fig.4h), respectively.}}
\vspace*{-0.5in}
\newpage
\begin{figure}
\subfigure[]{
\label{PictureThreeLabel}
\hspace*{-0.7 in}
\begin{minipage}[b]{0.5\textwidth}
\centering
\includegraphics[width=\textwidth]{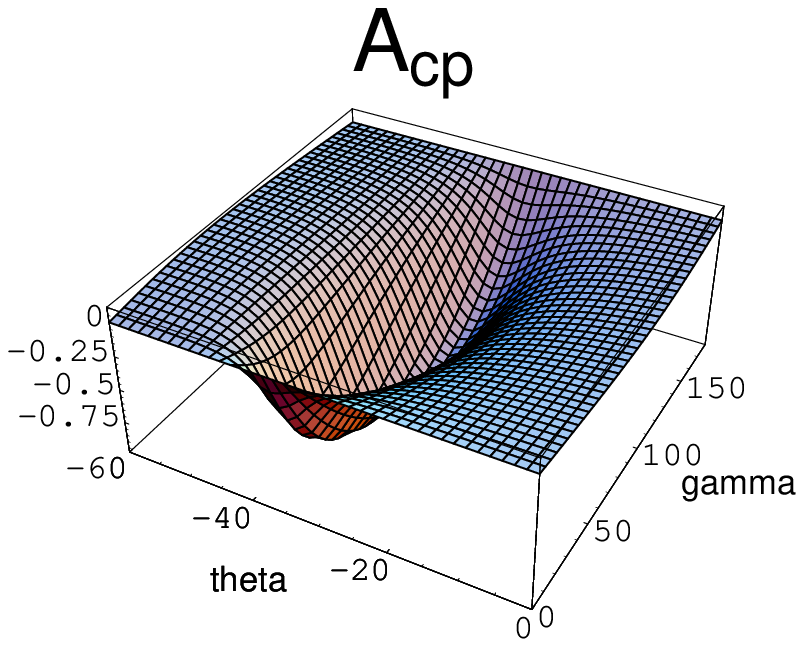}
\end{minipage}}
\subfigure[]{
\label{PictureFourLabel}
\hspace*{0.3in}
\begin{minipage}[b]{0.5\textwidth}
\centering
\includegraphics[width=\textwidth]{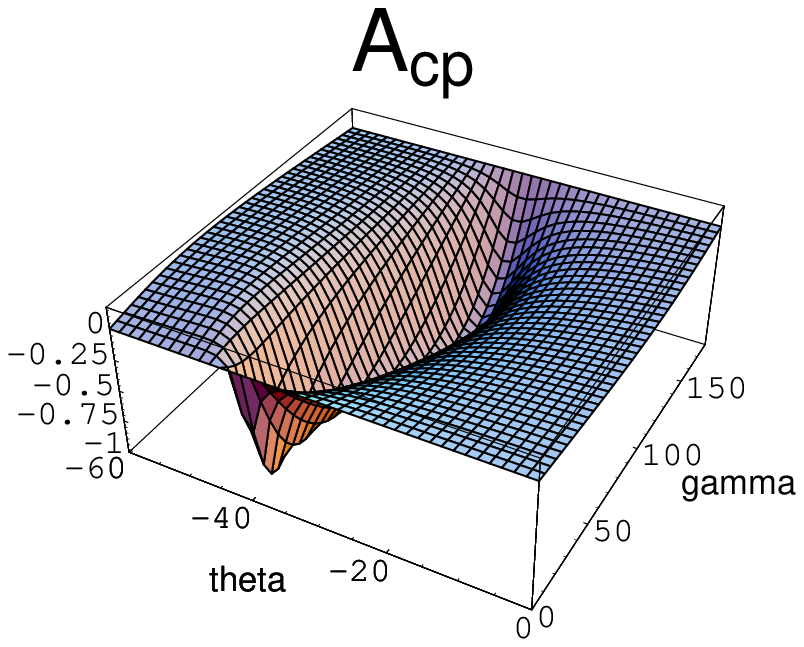}
\end{minipage}}
\end{figure}
\vspace*{-0.45in}
\noindent {\bf Figure 5[i,~j]}.
{{ \it Showing the CP-asymmetry $A_{cp}(B^+ \to \eta K^+)$ as a function 
of the mixing angle theta ($\theta$) and the unitarity angle gamma($\gamma$). 
 The angles are given in degree. The left one corresponds to the renormalization scale $\mu=m_b/2$ (Fig.5i), whereas the right one, $\mu = m_b$ (Fig.5j).}}
\vspace*{-0.5in}
\newpage
\begin{figure}
\subfigure[]{
\label{PictureThreeLabel}
\hspace*{-0.7 in}
\begin{minipage}[b]{0.5\textwidth}
\centering
\includegraphics[width=\textwidth]{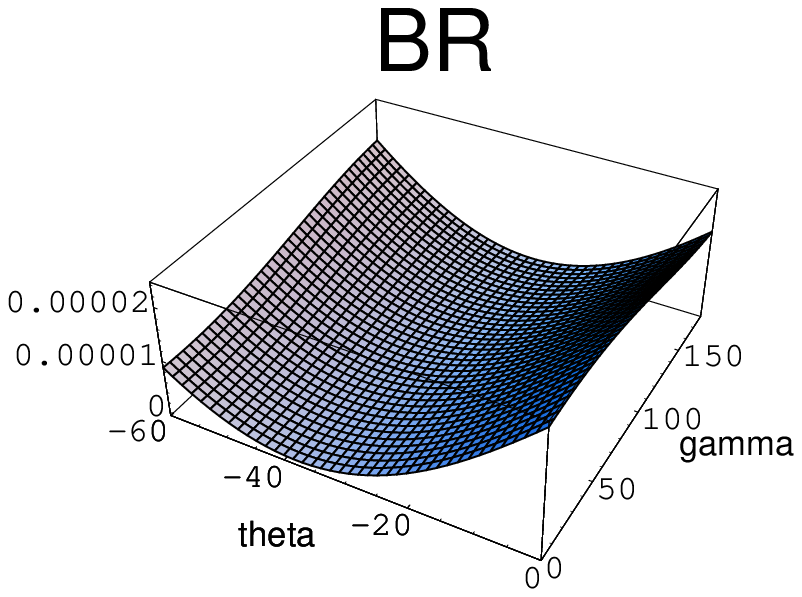}
\end{minipage}}
\subfigure[]{
\label{PictureFourLabel}
\hspace*{0.3in}
\begin{minipage}[b]{0.5\textwidth}
\centering
\includegraphics[width=\textwidth]{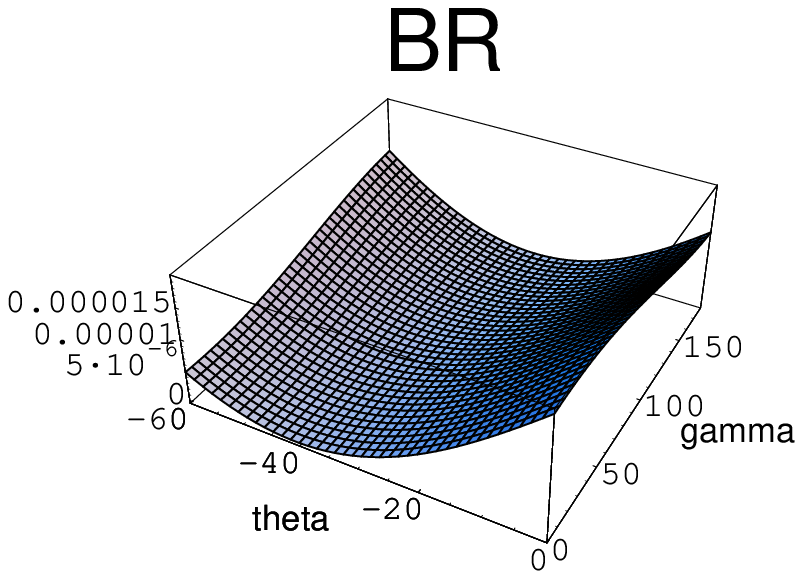}
\end{minipage}}
\end{figure}
\vspace*{-0.45in}
\noindent {\bf Figure 6[k,~l]}.
{{ \it Showing the $BR(B^+ \to \eta K^+)$ as a function 
of the mixing angle theta (degree) and the unitarity angle gamma (degree). The angles are in degree. The left one corresponds to the renormalization scale $\mu=m_b/2$ (Fig.6k), whereas, the right one,   $\mu = m_b$ (Fig.6l).}}
\end{document}